\documentclass[twocolumn,preprintnumbers,prc]{revtex4-1}

\usepackage{graphicx}
\usepackage{amssymb}
\usepackage{epstopdf}
\usepackage{bm}
\begin{document}
    
\title{A Comment on ``The possible explanation of neutron lifetime beam anomaly'' by A.~P.~Serebrov, {\em et al.}}
    
\author{F.~E.~Wietfeldt}
\affiliation{Department of Physics and Engineering Physics, Tulane University, New Orleans, LA 70118}
\author{R.~Biswas}
\affiliation{Department of Physics and Engineering Physics, Tulane University, New Orleans, LA 70118}
\author{R.~W.~Haun}\thanks{present address: Department of Physics, University of Maryland, College Park, MD 20742, USA}
\affiliation{Department of Physics and Engineering Physics, Tulane University, New Orleans, LA 70118}
\author{M.~S.~Dewey}
\affiliation{National Institute of Standards and Technology, Gaithersburg, MD 20899, USA}
\author{J.~Caylor}
\affiliation{Department of Physics, University of Tennessee, Knoxville, TN 37996, USA}
\author{N.~Fomin}
\affiliation{Department of Physics, University of Tennessee, Knoxville, TN 37996, USA}
\author{G.~L.~Greene}
\affiliation{Department of Physics, University of Tennessee, Knoxville, TN 37996, USA}
\affiliation{Physics Division, Oak Ridge National Laboratory, Oak Ridge, TN 37830, USA}
\author{C.~C.~Haddock}
\affiliation{National Institute of Standards and Technology, Gaithersburg, MD 20899, USA}
\author{S.~F.~Hoogerheide}
\affiliation{National Institute of Standards and Technology, Gaithersburg, MD 20899, USA}
\author{H.~P.~Mumm}
\affiliation{National Institute of Standards and Technology, Gaithersburg, MD 20899, USA}
\author{J.~S.~Nico}
\affiliation{National Institute of Standards and Technology, Gaithersburg, MD 20899, USA}
\author{B.~Crawford}
\affiliation{Physics Department, Gettysburg College, Gettysburg, PA 17325, USA}
\author{W.~M.~Snow}
\affiliation{Physics Department, Indiana University, Bloomington, IN 47405, USA}
\date{\today}

\maketitle

In a recent manuscript, Serebrov {\em et al.} \cite{Ser20} propose that loss of protons due to residual gas interactions in the most recent beam neutron lifetime experiment \cite{Nic05,Yue13} led to a systematic error that could account for the well known disagreement between the beam method \cite{Spi88,Byr96,Nic05,Yue13} and the ultracold neutron storage method \cite{Mam93,Ser05,Pic10,Ste12,Arz15,Ezh18,Pat18,Ser18}. In their paper, Serebrov {\em et al.} make a simplified model of the vacuum environment of the trap as a vessel with cold walls (the magnet bore) located inside another vessel with warm walls (the outer vacuum system). They assume that residual gas flows from the outer vessel into the inner vessel, remaining in gas phase at thermal equilibrium with the walls in the two vessels. Therefore the molecular density in the inner vessel reaches equilibrium at $n = P/k\sqrt{T_1 T_2}$, where $P$ is the vacuum pressure in the outer chamber, $k$ is the Boltzmann constant, and $T_1$, $T_2$ are the vessel temperatures. Using $P = 10^{-9}$ mbar as the ion gauge pressure (actually the upper limit as the gauge was under range) and $T_1 = 300$ K, $T_2 = 4$ K, they obtain $n = 2.1\times 10^8$ cm$^{-3}$ inside the trap. Later they show that, at such a density, charge exchange by trapped protons with residual gas components such as H$_2$O, CH$_4$, CO, and CO$_2$ would cause a significant loss during the 10 ms trap period and result in a measured neutron lifetime that is too long. While residual gas interactions should occur at some level, we find this analysis to be flawed because it neglects cryocondensation on the cold bore, a crucial feature of the trap vacuum.
\par
\begin{figure*}
\begin{center}
\includegraphics[width = 6in]{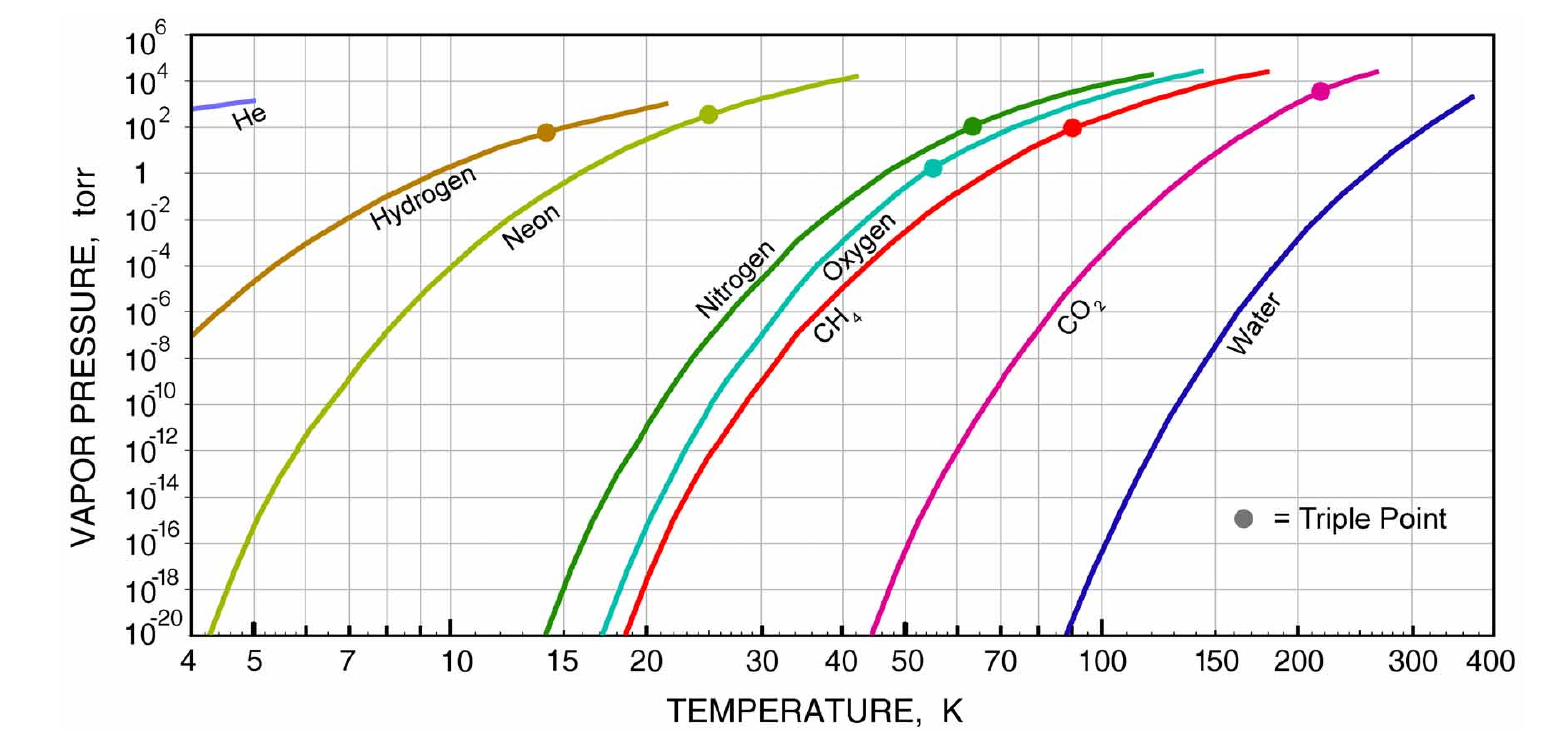}
\end{center}
\caption{\label{F:vaporP} Saturated vapor pressure of common gases as a function of temperature, from \cite{Ros19}.}
\end{figure*}

The cold bore of the magnet was a 45 cm long, 12 cm inner diameter stainless steel tube in direct contact with the liquid helium bath. Its operational temperature was about 8 K.  At this temperature the condensation coefficients of most gases are close to unity so residual gas will condense on the wall after just a few collisions, rather than remain in the gas phase and reach thermal equilibrium. The bore is effectively a cryopump. According to the theory of cryocondensation (see for example \cite{Bae87,Ohan}) the partial pressure of each gas component in the bore will reach equilibrium close to its saturation vapor pressure. Figure \ref{F:vaporP} shows a plot of saturation vapor pressure {\em vs.} temperature for a number of common gases. Other than hydrogen, helium, and neon the partial pressure and density of all residual gas components are predicted to be far lower than the estimate in \cite{Ser20}, although we note that determining the actual partial pressures of species inside the proton trap is a complicated problem that depends on many factors. There is no reason to expect neon in the vacuum system. One would expect hydrogen of course, and also helium due to its omnipresence in the guide hall atmosphere. Charge exchange with these species would result in trapped hydrogen (monatomic or diatomic) and helium ions that could be detected by the surface barrier detector after the trap is opened.

\par
In summary, we find the analysis of Serebrov {\em et al.} \cite{Ser20} to be incorrect due to their neglect of cryocondensation of residual gas on the cold magnet bore that encloses the proton trap in the beam lifetime experiment. More generally, for the past few years we have been actively investigating many systematic effects in the beam neutron lifetime experiment, including those that could be caused by residual gas and other vacuum related phenomena. This is primarily an experimental effort, as the apparatus is very complicated and difficult to model accurately and to useful precision in a simulation or calculation.


\begin{thebibliography}{99}
\bibitem{Ser20} A.~P.~Serebrov,  {\em et al.}, arXiv:2003.02092v1 [nucl-ex], (2020).
\bibitem{Nic05} J.~S.~Nico, {\em et al.}, Phys. Rev. C {\bf 71}, 055502 (2005).
\bibitem{Yue13} A.~T.~Yue, {\em et al.}, Phys. Rev. Lett. {\bf 111}, 222501 (2013).
\bibitem{Spi88} P.~E.~Spivak, JETP {\bf 67}, 1735 (1988).
\bibitem{Byr96} J.~Byrne, {\em et al.}, Europhys. Lett. {\bf 33}, 187 (1996).
\bibitem{Mam93} B.~Mampe, {em et al.}, JETP Lett. {\bf 57}, 82 (1993.)
\bibitem{Ser05} A.~Serebrov, {\em et al.}, Phys. Lett. B {\bf 605}, 72 (2005).
\bibitem{Pic10} A.~Pichlmaier, {\em et al.}, Phys. Lett. B {\bf 693}, 221 (2010).
\bibitem{Ste12} A.~Steyerl, {\em et al.}, Phys. Rev. C {\bf 85}, 065503 (2012).
\bibitem{Arz15} S.~Arzumanov, {\em et al.}, Phys.~Lett.~B {\bf 745}, 79 (2015).
\bibitem{Ezh18} V.~F.~Ezhov, {\em et al.}, JETP Lett. {\bf 107}, 671 (2018).
\bibitem{Pat18} R.~W.~Pattie, {\em et al.}, Science {\bf 360}, 627 (2018).
\bibitem{Ser18} A.~P.~Serebrov, {\em et al.}, Phys.~Rev.~C {\bf 97}, 055503 (2018).
\bibitem{Bae87} W.~G.~Baechler, {\em et al.}, Vacuum {\bf 37}, 21 (1987).
\bibitem{Ohan} J.~F.~O'Hanlon, {\em A User's Guide to Vacuum Technology}, 3/e, pp. 263--285, John Wiley \& Sons, USA (2003), ISBN 0-471-27052-0.
\bibitem{Ros19} R.~G.~Ross, in {\em Low Temperature Materials and Mechanisms}, ed. by Y.~Bar-Cohen, pp. 109-181, CRC Press, USA (2019), ISBN 0-367-87134-3.

\end{thebibliography}
\end{document}